\documentclass[]{mn2e}
\usepackage{graphicx}
\usepackage{epsfig}
\usepackage{amsmath}
\newcommand       \Angstrom     {\,{\rm \AA}}

\newcommand       \cm           {\,{\rm cm}}

\newcommand       \eV           {\,{\rm eV}}

\newcommand     \gtsim  {\lower.5ex\hbox{$\buildrel > \over \sim$}}
\newcommand     \ltsim  {\lower.5ex\hbox{$\buildrel < \over \sim$}}
\newcommand     \simgt  {\lower.5ex\hbox{$\buildrel > \over \sim$}}
\newcommand     \simlt  {\lower.5ex\hbox{$\buildrel < \over \sim$}}

\newcommand       \mum          {\,{\rm \mu m}}
\newcommand       \ppm          {\,{\rm ppm}}

\newcommand       \simali       {\sim\,}

\newcommand       \NH          {N_{\rm H}}
\newcommand       \NC          {N_{\rm C}}
\newcommand       \NCj          {N_{{\rm C},j}}
\newcommand       \cabsij      {C_{{\rm abs},j}^{+}}
\newcommand       \cabsaj     {C_{{\rm abs},j}^{-}}
\newcommand       \cabsnj     {C_{{\rm abs},j}^{0}}

\newcommand       \avgcabsiNC  {\langle C_{\rm abs}^{+}(\lambda)/\NC\rangle}
\newcommand       \avgcabsaNC  {\langle C_{\rm abs}^{-}(\lambda)/\NC\rangle}
\newcommand       \avgcabsnNC  {\langle C_{\rm abs}^{0}(\lambda)/\NC\rangle}

\newcommand       \avgcabsNC  {\langle C_{\rm abs}(\lambda)/\NC\rangle}
\newcommand       \CTOHPAH  {\left[{\rm C/H}\right]_{\rm PAH}}
\title{PAHs and the 2175$\Angstrom$ Extinction Bump}
\title{
Polycyclic Aromatic Hydrocarbon Molecules
and the 2175$\Angstrom$ Interstellar Extinction Bump
}
\author[Lin, Yang \& Li]
            {Qi~Lin$^{1,2}$,
              X.J.~Yang$^{1,2}$\thanks{xjyang@xtu.edu.cn},
             and Aigen Li$^{2}$\thanks{lia@missouri.edu}\\
 $^1$Hunan Key Laboratory for Stellar
              and Interstellar Physics
              and School of Physics and Optoelectronics,
              Xiangtan University, Hunan 411105, China\\
  $^2$Department of Physics and Astronomy,
                  University of Missouri,
                  Columbia, MO 65211, USA\\
                  }

\begin{document}

\date{}
\pagerange{\pageref{firstpage}--\pageref{lastpage}} \pubyear{2023}

\maketitle

\label{firstpage}
\begin{abstract}
The exact nature of the 2175$\Angstrom$
extinction bump, the strongest spectroscopic
absorption feature superimposed on the interstellar
extinction curve, remains unknown
ever since its discovery in 1965.
Popular candidate carriers for the extinction bump
include nano-sized graphitic grains
and polycyclic aromatic hydrocarbon (PAH) molecules.
To quantitatively evaluate PAHs as a possible carrier,
we perform quantum chemical computations
for the electronic transitions of 30 compact,
pericondensed PAH molecules
and their cations as well as anions
with a wide range of sizes from 16 to 96 C atoms
and a mean size of 43 C atoms.
It is found that a mixture of such PAHs,
which individually exhibit sharp absorption features,
show a smooth and broad absorption band
that resembles the 2175$\Angstrom$
interstellar extinction bump.
Arising from $\pi^{\ast}$\,$\leftarrow$\,$\pi$ transitions,
the width and intensity of the absorption bump
for otherwise randomly-selected
and uniformly-weighted PAH mixtures,
do not vary much with PAH sizes and charge states,
whereas the position shifts to longer wavelengths
as PAH size increases.
While the computed bump position,
with the computational uncertainty
taken into account, appears to agree with
that of the interstellar extinction bump,
the computed width is considerably broader
than the interstellar bump
{\it if} the molecules are uniformly weighted.
It appears that, to account for the observed
bump width, one has to resort to PAH species 
of specific sizes and structures.
\end{abstract}

\section{Introduction}\label{sec:intro}
The 2175$\Angstrom$ extinction bump,
the most prominent spectral feature
superimposed on the interstellar extinction curve
spanning roughly the wavelength range
between 1700 and 2700$\Angstrom$,
has puzzled astronomers for nearly six decades
ever since its first detection (Stecher 1965).
It is widely seen in the Milky Way
(e.g., see Fitzpatrick \& Massa 1986, 2007;
Valencic et al.\ 2004;
but also see Clayton et al.\ 2000)
and nearby galaxies,
including the Large Magellanic Cloud
(e.g., see Fitzpatrick 1986,
Misselt et al.\ 1999,
Gordon et al.\ 2003),
several regions in the Small Magellanic Cloud
(Gordon \& Clayton 1998,
Ma\'iz-Apell\'aniz \& Rubio 2012,
Gordon et al.\ 2003)
and M31 (Bianchi et al.\ 1996, Dong et al.\ 2014,
Clayton et al.\ 2015, Wang et al.\ 2022).
It has also been seen in more distant galaxies.
Motta et al.\ (2002) reported the detection
of the 2175$\Angstrom$ extinction bump
in the lens galaxy of the gravitational lens
system SBS\,0909+532 at $z\approx0.83$.
%
York et al.\ (2006) detected the 2175$\Angstrom$
bump in a damped Ly$\alpha$ absorber (DLA)
at $z\approx0.524$ toward AO\,0235+164,
a quasar at $z=0.94$.
%
%
This extinction bump has also been seen
in over a dozen of dusty intervening MgII
systems at $z$\,$\approx$1--2 toward quasars
in the Sloan Digital Sky Survey (SDSS) data base
(see Wang et al.\ 2004, Srianand et al.\ 2008,
Zhou et al.\ 2010, Jiang et al.\ 2011,
Ma et al.\ 2015, 2017).
El{\'{\i}}asd{\'o}ttir et al.\ (2009) found
the 2175$\Angstrom$ bump in the optical afterglow
spectrum of GRB~070802, a gamma-ray burst (GRB)
at a redshift of $z\approx2.45$.
Several other GRBs also show evidence
for the presence of this bump
(see Liang \& Li 2009, 2010;
Prochaska et al.\ 2009;
Zafar et al.\ 2011, 2012).
%
%
%
The attenuation of starlight by interstellar dust
in star-forming galaxies at $1 < z < 3$
has also indicated the possible presence of
the 2175$\Angstrom$ bump
(e.g., see Noll et al.\ 2007, 2009;
Conroy et al.\ 2010;
Kriek et al.\ 2013;
Battisti et al.\ 2020;
Shivaei et al.\ 2022).
Very recently, the 2175$\Angstrom$ extinction
bump was detected by
the {\it James Webb Space Telescope} (JWST)
in a distant galaxy at redshift $z\approx6.71$
(Witstok et al.\ 2023).
%


Despite nearly 60 years' extensive observational,
theoretical and experimental studies, the exact
carrier of the 2175$\Angstrom$ extinction bump
remains unidentified.
Stecher \& Donn (1965) first proposed small graphite
grains as a carrier of the extinction bump,
but decades of observations led to discrediting
this hypothesis since it cannot account for an
important observational constraint, i.e.,
the central wavelength of the bump is virtually
invariant across a large number of sightlines
while its width can vary significantly
(Fitzpatrick \& Massa 1986; Mathis 1994).
When increasing the size of graphite grains,
the width does grow larger, but at the expense
of the peak shifting to longer wavelengths
(e.g., see Draine \& Malhotra 1993).

A popular hypothesis attributes the 2175$\Angstrom$
bump to polycyclic aromatic hydrocarbon (PAH) molecules
(Joblin et al.\ 1992; Li \& Draine 2001;
Malloci et al.\ 2004, 2008;
Cecchi-Pestellini et al.\ 2008;
Steglich et al. 2010, 2012)
which reveal their presence in the interstellar medium
(ISM) through a distinct series of emission bands
in the mid infrared (IR) at 3.3, 6.2, 7.7, 8.6 and 11.3$\mum$.
Massa et al.\ (2022) recently analyzed the low-resolution
$\simali$5--14$\mum$ spectra of the diffuse PAH emission
obtained by the {\it Infrared Spectrograph} (IRS)
on board the {\it Spitzer Space Telescope}
toward a carefully selected sample of 16 stars.
The sightlines toward these stars have well-determined
ultraviolet (UV) extinction curves.
They found a strong correlation between the strength of
the 2175$\Angstrom$ bump and the PAH emission bands,
supporting PAHs as a possible carrier of the bump.
%

%

To quantitatively evaluate PAHs as a possible carrier
of the 2175$\Angstrom$ extinction bump,
we perform quantum chemical computations
for the electronic transitions
of 30 compact, pericondensed PAH molecules
and their cations as well as anions.
These molecules span a wide range of sizes
from 16 to 96 carbon (C) atoms
and have a mean size of 43 C atoms.
We analyze the mean spectra of these molecules
and compare them with the interstellar
2175$\Angstrom$ bump.
This paper is organized as follows.
In \S\ref{sec:methods} we briefly describe
the computational methods and target molecules.
The computed spectra are presented
and discussed in \S\ref{sec:results}.
Our major conclusion is summarized
in \S\ref{sec:summary}.

\section{Computational Methods and
            Target Molecules}\label{sec:methods}
We use the Time-Dependent Density Functional
Theory (TD-DFT) as implemented in the software
{\sf Octopus} to calculate the UV/visible electronic
spectra of a series of PAHs in anion, neutral, and
cationic charging states (Marques et al.\ 2003).
We employ the hybrid functional B3LYP
in combination with the following crucial parameters:
(1) the size of numerical box (each atom coated by
a sphere of radius 3$\Angstrom$ for neutrals
and cations, 5$\Angstrom$ for anions),
(2) the grid spacing of 0.3$\Angstrom$,
(3) the time integration length of 20\,$\hbar\eV^{-1}$,
and (4) the time step of 0.002\,$\hbar\eV^{-1}$.
Before computing the electronic spectra,
we perform structure optimization
with the software {\sf Gaussian09}
by using the B3LYP functional
in conjunction with the 6-31+G(d) basis
set (Becke 1993; Frish et al.\ 1984).
As catacondensed PAHs with open structures
are less stable in the ISM, we select 30 compact,
pericondensed PAHs (see Figure~\ref{fig:target}).
These molecules span a wide range of sizes
from pyrene with 16 C atoms (i.e., $\NC=16$)
to circumcircumcoronene and circumcircumpyrene
with $\NC=96$.
Starting from pyrene (C$_{16}$H$_{10}$),
perylene (C$_{20}$H$_{12}$),
anthracene (C$_{22}$H$_{12}$),
coronene (C$_{24}$H$_{12}$),
and ovalene (C$_{32}$H$_{16}$)
as seed molecules,
we investigate the effects of size and charging
on the electronic absorption spectra of large,
compact PAHs up to circumcircumcoronene
and circumcircumpyrene.
In the ISM, typical PAHs have $\NC\approx100$
(see Li \& Draine 2001). However, it is computationally
not feasible for TD-DFT to consider PAHs
much larger than circumcircumcoronene
or circumcircumpyrene.

%
%

It is interesting to note that apart from being stable,
these PAH molecules are anticipated to serve as
the primary predecessors for the formation of soot
in hydrocarbon flames
(Frenklach \& Feigelson 1989;
Cherchneff \& Barker 1992).
We also remark that these molecules
have been adopted by Croiset et al.\ (2016)
to map the sizes of PAHs in NGC~7023,
a reflection nebula, using data from
the {\it Stratospheric Observatory
for Infrared Astronomy} (SOFIA).
%

\section{Results and Discussion}\label{sec:results}
Let $\cabsij(\lambda)/\NC$, $\cabsaj(\lambda)/\NC$,
and $\cabsnj(\lambda)/\NC$ be the absorption cross
sections per C atom of the $j$-th PAH cation, anion
and neutral species at wavelength $\lambda$.
Let $\NCj$ be the number of C atoms of
the $j$-th PAH species.
The mean absorption cross sections
(per C atom) for the 30 PAH cations,
anions and neutrals considered in this work
are respectively
\begin{equation}\label{eq:avgcabsi}
\avgcabsiNC = \frac{1}{30} \sum_{j=1}^{30} \cabsij(\lambda)/\NCj ~,
\end{equation}
\begin{equation}\label{eq:avgcabsi}
\avgcabsaNC = \frac{1}{30} \sum_{j=1}^{30} \cabsaj(\lambda)/\NCj ~,
\end{equation}
\begin{equation}\label{eq:avgcabsi}
\avgcabsnNC = \frac{1}{30} \sum_{j=1}^{30} \cabsnj(\lambda)/\NCj ~.
\end{equation}
The overall mean absorption cross sections
(per C atom) are obtained from averaging
over that of all three charge states
(and assuming equal weights for cations,
anions and neutrals):
\begin{equation}\label{eq:avgcabsi}
\avgcabsNC = \frac{1}{3} \avgcabsiNC
+ \frac{1}{3} \avgcabsnNC
+ \frac{1}{3} \avgcabsaNC ~~.
\end{equation}
%

\begin{figure*}
\vspace{-4mm}
\begin{center}
\includegraphics[width=10cm,angle=0]{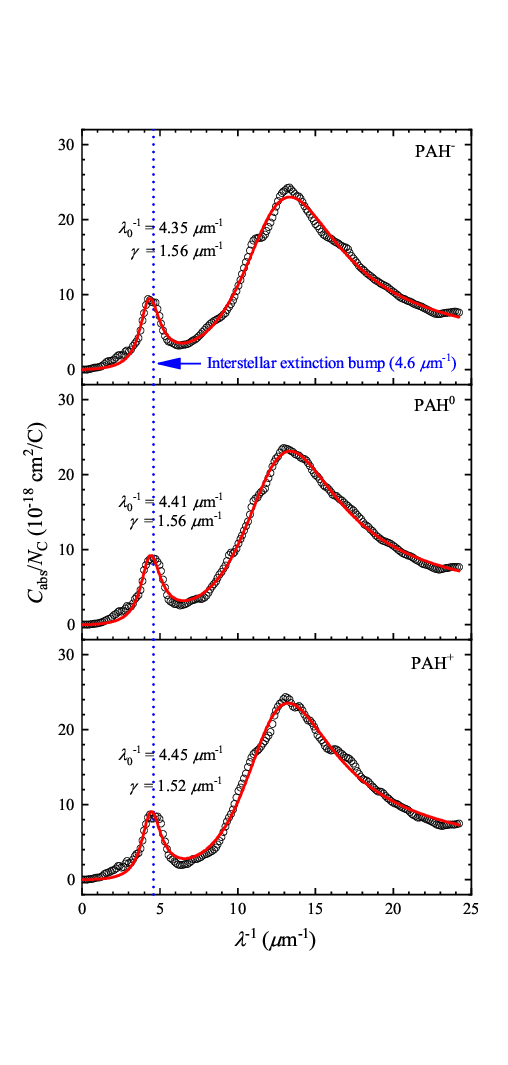}
\end{center}
\vspace{-5mm}
\caption{\label{fig:cabs_ion_neu} 
  Mean absorption spectra (black open circles)
  of a mixture of 30 compact PAH anions (upper panel),
  neutrals (middle panel) and cations (bottom panel).
  Also shown are the spectra
  fitted with a Drude profile
  for the $\pi^{\ast}$\,$\leftarrow$\,$\pi$ transitions
  and a Fano profile for the transitions
  involving $\sigma$ electrons (red solid lines).
  The vertical blue dashed line shows
  the peak of the 2175$\Angstrom$ ($4.6\mum^{-1}$)
  interstellar extinction bump.
  }
\vspace{-5mm}
\end{figure*}

%
Figure~\ref{fig:cabs_ion_neu} shows $\avgcabsiNC$,
$\avgcabsaNC$ and $\avgcabsnNC$, the mean absorption
spectra for the target molecules in different charge states,
weighted on a per C atom basis.
It is apparent that, although a single PAH species
exhibits sharp, individual absorption features,
the mean spectrum of a mixture of many individual
molecules, with many overlapping absorption features,
effectively smooth out these sharp features
and produce two prominent broad absorption bands
in the UV, respectively arising from transitions
involving $\pi$ and $\sigma$ electrons
(see Li \& Draine 2001).
This has already been demonstrated by
previous experimental
(Joblin et al.\ 1992;
Steglich et al. 2010, 2012)
and quantum computational studies
(Malloci et al.\ 2004, 2008;
Cecchi-Pestellini et al.\ 2008).
Figure~\ref{fig:cabs_ion_neu}
confirms these earlier studies
that, indeed, the sharp absorption peaks
of individual species are smoothed out
in PAH mixtures, and a concentration of
strong absorption features
in the 2000--2400\AA\ wavelength region
blend together to produce a pronounced
band resembling the 2175$\Angstrom$
extinction bump. In addition, an even more
prominent absorption band is seen
around 12$\mum^{-1}$.

As illustrated in Figure~\ref{fig:cabs_ion_neu},
the low-lying band at $\simali$4.4$\mum^{-1}$
is dominated by strong
$\pi^{\ast}$\,$\leftarrow$\,$\pi$ transitions,
while the broader band at $\simali$12$\mum^{-1} $
is caused by the sum of
$\sigma^{\ast}$\,$\leftarrow$\,$\sigma$,
$\sigma^{\ast}$\,$\leftarrow$\,$\pi$,
$\pi^{\ast}$\,$\leftarrow$\,$\sigma$,
Rydberg transitions and possible plasmon effects.
Its low-energy tail may contribute to the far-UV
extinction rise at $\lambda^{-1}>5.9\mum^{-1}$
in the interstellar extinction curve,
particularly the so-called ``nonlinear''
far-UV rise (see eq.\,1 of Fitzpatrick \& Massa 1988,
eqs.\,4a,b of Cardelli et al.\ 1989).
 %


To quantitatively evaluate PAHs as a possible
carrier of the 2175$\Angstrom$ extinction bump,
we construct a phenomenological model to
characterize the mean absorption spectra of
PAH cations, anions and neutrals in terms of
a Drude profile of width $\gamma_1$ and
peak position of $x_{0,1}$
for the $\pi^{\ast}$\,$\leftarrow$\,$\pi$ transitions,
and an asymmetrical Fano profile
of width $\gamma_2$ and
peak position of $x_{0,2}$
for the collective electronic transitions arising from
$\sigma^{\ast}$\,$\leftarrow$\,$\sigma$,
$\sigma^{\ast}$\,$\leftarrow$\,$\pi$,
$\pi^{\ast}$\,$\leftarrow$\,$\sigma$,
and Rydberg transitions as well as possible
plasmon effects:
\begin{equation}\label{eq:fit}
C_{\mathrm{abs} }(\lambda) /N_{\mathrm{C} }
= a_{1} D\left (x;x_{0,1},\gamma_1\right)
+ a_{2} F\left (x;x_{0,2},\gamma_2 \right) ~~,
\end{equation}
where $x\equiv1/\lambda$ is the inverse
wavelength, $D\left (x;x_{0,1},\gamma_1\right)$
is the Drude function, $x_{0,1}$ and $\gamma_1$
are respectively its peak and width:
\begin{equation}\label{eq:drude}
D\left (x;x_{0,1},\gamma \right )
= \frac{x^{2} }{\left(x^{2}-x_{0,1}^{2}\right )^{2}
  +x^{2}\gamma _{1}^{2}} ~~,
\end{equation}
and the Fano function is
\begin{equation}\label{eq:fano}
F\left(x;x_{0,2},\gamma_2\right )
=\frac{\left[\left(x-x_{0,2}\right )
+q\gamma _{2}/2 \right ]^{2}}
{\left (x-x_{0,2}\right )^{2}
+\left (\gamma _{2}/2\right )^{2}} ~~,
\end{equation}
where $q$ is a dimensionless,
phenomenological parameter for depicting
the line shape asymmetry (see Fano 1961).

As shown in Figure~\ref{fig:cabs_ion_neu},
the mean absorption spectra of PAH cations,
anions and neutrals arising from
$\pi^{\ast}$\,$\leftarrow$\,$\pi$ transitions
are closely fitted by the combination of
a Drude profile and a Fano profile.
The parameters are tabulated in Table~\ref{tab:para}.
The absorption spectra of PAHs of all
three charge states all peak around
4.4$\mum$, differing only by $\simali$1\%.
The widths of the absorption spectra of all
three charge states are all around 1.56$\mum^{-1}$,
differing only by $\simlt$2\%.
The intensities of $\pi^{\ast}$\,$\leftarrow$\,$\pi$
transitions are also very similar for all the charge states,
differing only by $\simlt$5\%.
This contradicts the experimental findings
that, upon ionization, the absorption bump due to
$\pi^{\ast}$\,$\leftarrow$\,$\pi$ transitions
substantially weakens
(see Lee \& Wdowiak 1993, Robinson et al.\ 1997).

\begin{figure*}
\vspace{-3mm}
\begin{center}
\includegraphics[width=15cm,angle=0]{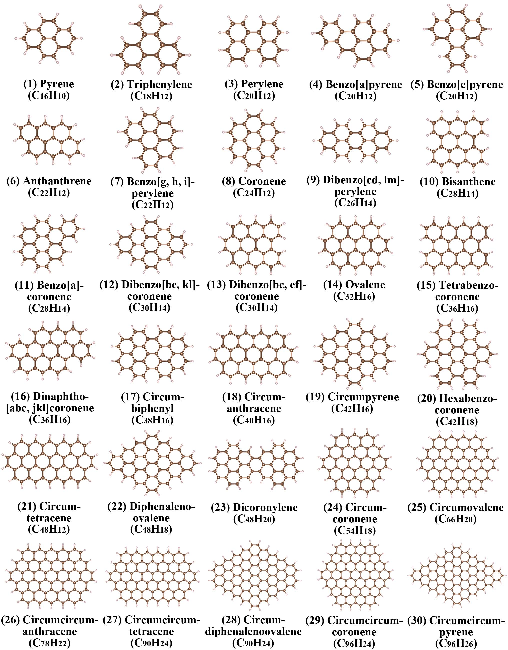}
\end{center}
\vspace{-5mm}
\caption{\label{fig:target} 
  Target molecules for TD-DFT computations
  of their electronic transitions.
  }
\end{figure*}

\begin{figure*}
\vspace{-3mm}
\begin{center}
\includegraphics[width=14.2cm,angle=0]{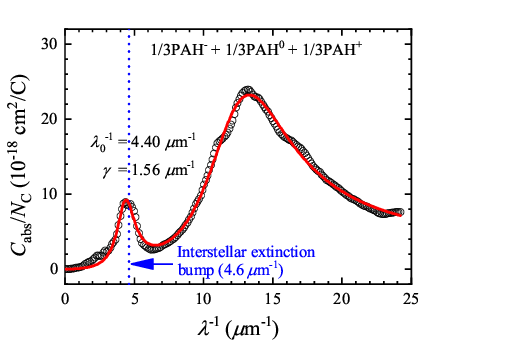}
\end{center}
\vspace{-5mm}
\caption{\label{fig:cabs_mean} 
  Mean absorption spectrum (black open circles)
  obtained by averaging over the mean spectra of
  PAH cations, anions and neutrals
  shown in Figure~\ref{fig:cabs_ion_neu}.
  Also shown is the spectrum
  fitted with a Drude profile
  and a Fano profile (red solid line).
  The vertical blue dashed line shows
  the peak of the interstellar extinction bump.
	 }
\vspace{-3mm}
\end{figure*}

Figure~\ref{fig:cabs_mean} shows $\avgcabsNC$,
the overall mean absorption spectrum of PAHs,
obtained by averaging over $\avgcabsiNC$,
$\avgcabsaNC$ and $\avgcabsnNC$,
the mean absorption spectra
for PAH cations, anions and neutrals
(on a per C atom basis).
We assign equal weights to all charge states.
We also fit $\avgcabsNC$ with a Drude profile
and a Fano profile.
The band strength of the $\pi^{\ast}$\,$\leftarrow$\,$\pi$
transitions for the mixture of those 30 compact PAHs
shown in Figure~\ref{fig:target} is
\begin{equation}\label{eq:cabs_int}
\int_{\pi^{\ast}\,\leftarrow\,\pi}
\Delta \avgcabsNC\,d\lambda^{-1}
\approx 2.05\times10^{-13}\cm/{\rm C} ~~,
\end{equation}
where $\Delta \avgcabsNC$ is the {\it excess}
mean absorption cross section of PAHs
at the $\pi^{\ast}$\,$\leftarrow$\,$\pi$ transitions.
Observationally, in the diffuse ISM
the optical depth of the 2175$\Angstrom$
extinction bump per hydrogen column
integrated over the bump
in inverse wavelength is
\begin{equation}\label{eq:cabs_int}
\int_{2175\Angstrom}
\Delta \tau_{2175}/\NH\,d\lambda^{-1}
\approx 8.2\times10^{-18}\cm/{\rm H} ~~,
\end{equation}
where $\Delta \tau_{2175}$ is the excess
optical depth at  2175$\Angstrom$,
and $\NH$ is the hydrogen column density
(see Table 1 of Draine 1994).
To account for the observed 2175$\Angstrom$
extinction bump in the diffuse ISM,
the amount of C (relative to H)
required to be tied up in PAHs is
\begin{equation}\label{eq:C2H}
\CTOHPAH =
\frac{\int_{2175\Angstrom}
\Delta\tau_{2175}/\NH\,d\lambda^{-1}}
{\int_{\pi^{\ast}\,\leftarrow\,\pi}
\Delta \avgcabsNC\,d\lambda^{-1}}
\approx 40\ppm ~~,
\end{equation}
where ppm is parts per million.

\begin{figure*}
\vspace{-3mm}
\begin{center}
\includegraphics[width=12cm,angle=0]{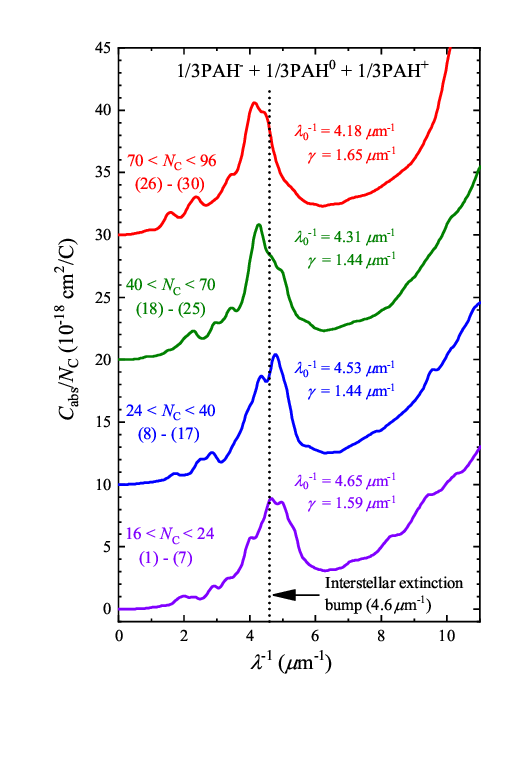}
\end{center}
\vspace{-5mm}
\caption{\label{fig:cabs_size} 
  Mean absorption spectra
  obtained by averaging over the mean spectra of
  PAH cations, anions and neutrals
  of different size ranges.
  	 }
\vspace{-3mm}
\end{figure*}

As shown in Figure~\ref{fig:cabs_mean},
the $\pi^{\ast}$\,$\leftarrow$\,$\pi$
absorption bump peaks at $\simali$4.4$\mum^{-1}$
and has a width of $\gamma\approx1.56\mum^{-1}$.
While it resembles the interstellar extinction bump
at 2175$\Angstrom$ or 4.6$\mum^{-1}$,
it peaks at a somewhat longer wavelength.
Apparently, this is not an charge effect since,
as shown in Figure~\ref{fig:cabs_ion_neu},
the peak wavelengths are very similar for
PAH cations, anions and neutrals.
To explore if this is a PAH size effect,
we divide our sample of 30 molecules
into four size groups: $16<\NC<24$,
$24<\NC<40$, $40<\NC<70$,
and $70<\NC<96$. For each size group,
we obtain the mean absorption spectrum
derived by averaging over all sizes and
charge states and show in Figure~\ref{fig:cabs_size}.
Clearly, as PAH sizes increase,
the $\pi^{\ast}$\,$\leftarrow$\,$\pi$ absorption bump
shifts to longer wavelengths.
Except for the smallest size group,
all peak at wavelengths somewhat longer than
the extinction bump. The $16<\NC<24$ group
exhibits a bump with a wavelength somewhat
shortward of the interstellar extinction bump.
However, these molecules are in the smallest
end in the PAH size distribution and are not
expected to dominate the contribution to
the extinction bump. Therefore, the mismatch
in the bump peak position is not a PAH size effect.

\begin{figure*}
\vspace{-3mm}
\begin{center}
\includegraphics[width=10cm,angle=0]{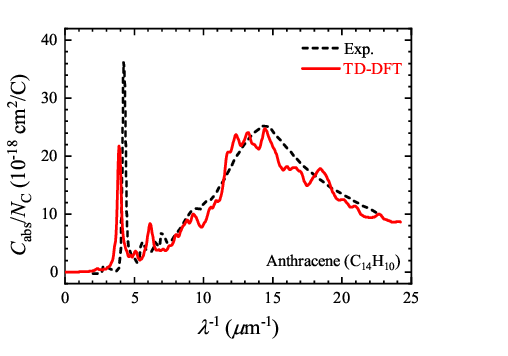}
\end{center}
\vspace{-5mm}
\caption{\label{fig:dft_exp} 
  Comparison of the computed absorption
  spectrum of anthracene (C$_{14}$H$_{10}$; red solid line)
  with the experimental gas-phase spectrum
  (black dashed line).
  	 }
\vspace{-3mm}
\end{figure*}

It is likely that the mismatch in position between
the computed $\pi^{\ast}$\,$\leftarrow$\,$\pi$
absorption bump and the interstellar extinction
bump is due to computational uncertainty.
Indeed, as illustrated in Figure~\ref{fig:dft_exp},
the DFT-computed absorption spectrum of
anthracene deviates from the experimental
spectrum of gas-phase anthracene
measured by Joblin (1992) by $\simali$0.4$\eV$:
while the $\pi^{\ast}$\,$\leftarrow$\,$\pi$ bump
in the gas-phase experimental spectrum
peaks at $\simali$4.24$\mum^{-1}$,
the DFT spectrum peaks at a longer wavelength
of $\simali$3.91$\mum^{-1}$.
As a matter of fact, the uncertainty for the band
positions at the B3LYP/6-31+G(d) level of TD-DFT
is $\simali$0.3$\eV$ (e.g., see Hirata et al.\ 1999).
With this uncertainty taken into account,
the $\pi^{\ast}$\,$\leftarrow$\,$\pi$ bump
for the mean absorption spectrum $\avgcabsNC$
of all species and all charge states would shift
from $\simali$4.40$\mum^{-1}$
to $\simali$4.64$\mum^{-1}$, concurring with
the interstellar extinction bump.

Unlike the central wavelength of
the $2175\Angstrom$ extinction bump
which is remarkably stable and
independent with environment,
the width shows considerable variation
and environmental dependence.
Fitzpatrick \& Massa (1986) found
a wide range of widths
($0.77\simlt\gamma\simlt1.25\mum^{-1}$)
for a sample of 45 stars, with an average
width of $\gamma\approx0.99\mum^{-1}$.
Valencic et al.\ (2004) found a wider range of
widths ($0.63\simlt\gamma\simlt1.47\mum^{-1}$)
for a sample of 417 stars, with an average
width of $\gamma\approx0.92\pm0.12\mum^{-1}$.
Even in the sample of Valencic et al.\ (2004),
very broad bumps are rare, as only 21 out of 417
sight lines had $\gamma>1.1\mum^{-1}$.
However, the widths of the DFT-computed
absorption spectra of PAH mixtures are all
larger than 1.4$\mum^{-1}$
(see Figures~\ref{fig:cabs_ion_neu}--\ref{fig:cabs_size}).
Apparently, the width is not an effect of PAH
charge states since, as shown in
Figure~\ref{fig:cabs_ion_neu},
the widths are rather similar
for different charge states.
On the other hand,
the width is not an effect of PAH size either
since, as shown in Figure~\ref{fig:cabs_size},
the widths of PAHs of different sizes
do not show any systematic variations.

\begin{figure*}
\vspace{-3mm}
\begin{center}
\includegraphics[width=10cm,angle=0]{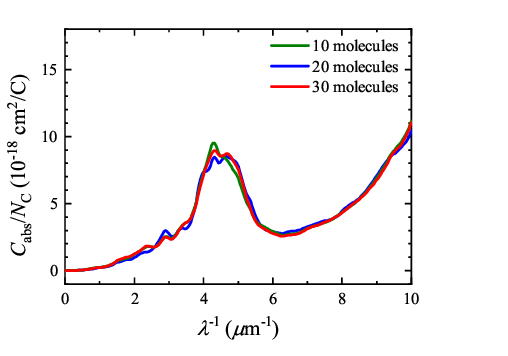}
\end{center}
\vspace{-5mm}
\caption{\label{fig:Nmolecules} 
  Comparison of the mean absorption spectra
  obtained for 10 and 20 molecules radnomly
  selected from those shown in Figure~\ref{fig:target}
  with that for a complete sample of 30 molecules.
  	 }
\vspace{-3mm}
\end{figure*}

It is speculated that the more molecules are considered,
the broader the $\pi^{\ast}$\,$\leftarrow$\,$\pi$ bump
would be. To examine this, we obtain the mean absorption
spectra of 10 and 20 molecules, randomly selected among
our sample of 30 molecules.
As shown in Figure~\ref{fig:Nmolecules},
the widths of the $\pi^{\ast}$\,$\leftarrow$\,$\pi$
absorption bump for a mixture of 10 and 20 molecules
are essentially the same as that for a mixture of 30 molecules.
This indicates that the bump width is not sensitive to
the exact number of molecules.

\begin{figure*}
\vspace{-3mm}
\begin{center}
\includegraphics[width=10cm,angle=0]{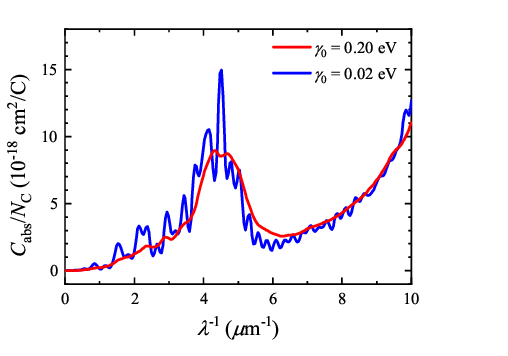}
\end{center}
\vspace{-5mm}
\caption{\label{fig:dft_width} 
  Comparison of the mean absorption spectrum
  obtained by assigning a width of $\gamma_0=0.2\eV$
  for each individual electronic transition (red line)
  with that of $\gamma_0=0.02\eV$ (blue line).
  	 }
\vspace{-3mm}
\end{figure*}

It is also speculated that the $\pi^{\ast}$\,$\leftarrow$\,$\pi$
bump width may vary with the width assigned to
each electronic transition. In computing the absorption
spectra of PAHs, we assume a width of
$\gamma_0=0.2\eV$ for each electronic transition.
To explore the effects of the assumed width of
each electronic transition ($\gamma_0$),
we consider $\gamma_0=0.02\eV$ and
compare in Figure~\ref{fig:dft_width}
the mean absorption spectra of PAHs
obtained with $\gamma_0=0.02\eV$
with that obtained with $\gamma_0=0.2\eV$.
It is apparent that the $\pi^{\ast}$\,$\leftarrow$\,$\pi$
bump width is not appreciably affected
by $\gamma_0$, the width assigned to
each electronic transition,
except that the spectrum with $\gamma_0=0.02\eV$
exhibits many fine structures. 
Therefore, we admit that the interstellar extinction
bump width remains unaccounted for
by the PAH mixtures considered here,
if they are {\it uniformly} weighted.

\begin{figure*}
\vspace{-3mm}
\begin{center}
\includegraphics[width=22cm,angle=90]{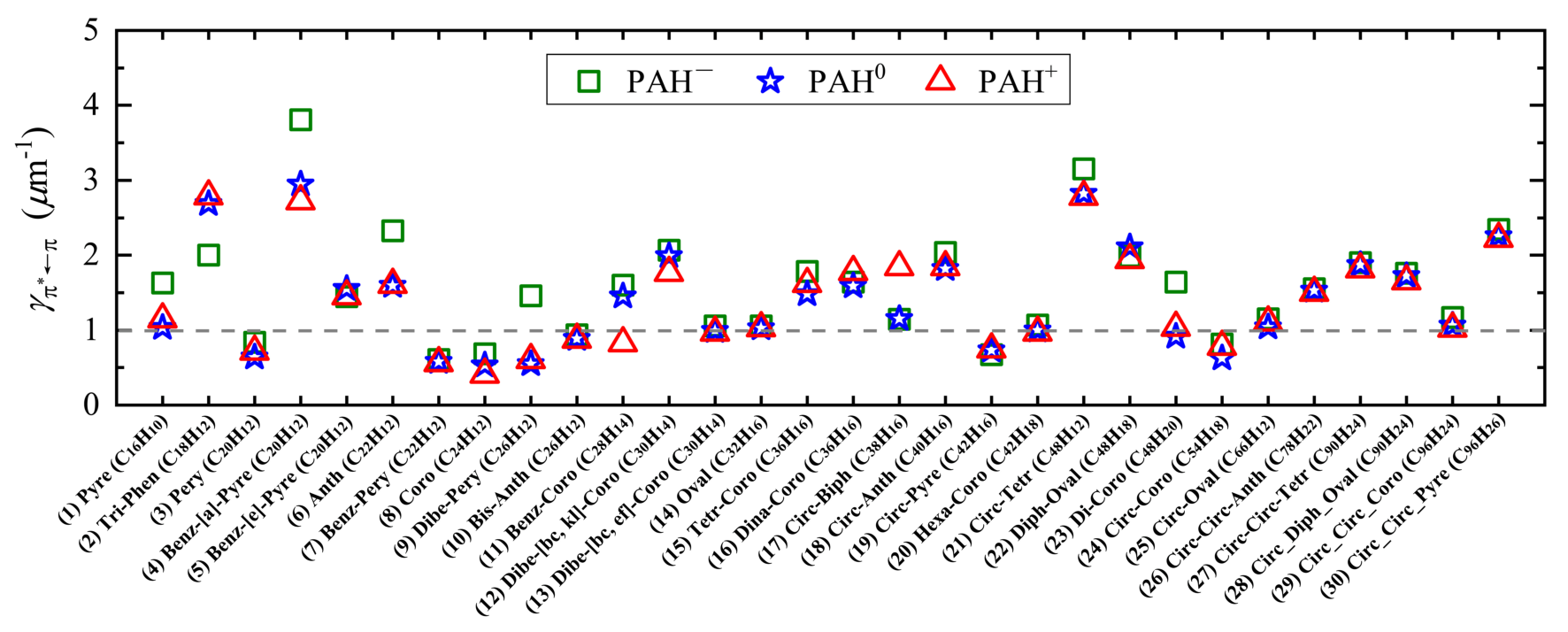}
\end{center}
\vspace{-5mm}
\caption{\label{fig:specpah} 
  The widths of the $\pi^{\ast}$\,$\leftarrow$\,$\pi$
  transitions of individual PAH species
  of three charge states
  (neutrals: blue stars; cations: red triangles;
  anions: green squares).
  We assign a width of  $\gamma_0=0.2\eV$
  for each individual electronic transition.
  }
\vspace{-3mm}
\end{figure*}

However, we should also note that this work
is limited to 30 idealized PAH species, most of
which are highly symmetric and undefective.
While our results suggest that when {\it uniformly}
weighted, such ideal PAHs are ruled out as being
responsible {\it alone} for the interstellar extinction bump,
it is difficult to disentangle accurate effects
for specific PAHs. Therefore, we examine
in Figure~\ref{fig:specpah} the width of
the $\pi^{\ast}$\,$\leftarrow$\,$\pi$ transition
of each PAH species, each in three charge states,
with a width of $\gamma_0=0.2\eV$
assumed for each electronic transition.
Figure~\ref{fig:specpah} demonstrates that
the bump widths are essentially the same
for all PAH neutrals and cations
except circumbiphenyl (C$_{38}$H$_{16}$;
$\#17$ in Figure~\ref{fig:target}).
In contrast, the bump widths of five (of 30)
PAH anions are appreciably broader than
that of neutrals and cations, and one PAH
anion (triphenylene, C$_{18}$H$_{12}$;
$\#2$ in Figure~\ref{fig:target}) shows
a narrower bump than its neutral and
cationic counterparts.

Figure~\ref{fig:specpah} also clearly
shows that the bump width varies
considerably from one species to another,
with 11 (of 30) species having a narrow width
of $\gamma_{\pi^{\ast}\leftarrow\pi}\simlt1.0\mum^{-1}$.
More specifically, small- and medium-sized,
highly symmetric species exhibit a narrow bump
(e.g., $\gamma_{\pi^{\ast}\leftarrow\pi}\approx0.53\mum^{-1}$
for coronene C$_{24}$H$_{12}$,
$\gamma_{\pi^{\ast}\leftarrow\pi}\approx0.73\mum^{-1}$
for circumpyrene C$_{42}$H$_{16}$,
and $\gamma_{\pi^{\ast}\leftarrow\pi}\approx0.62\mum^{-1}$
for circumcoronene C$_{54}$H$_{18}$).
In addition, several small PAHs with a less compact
structure also show narrow bumps
(e.g., $\gamma_{\pi^{\ast}\leftarrow\pi}\approx0.64\mum^{-1}$
for perylene C$_{20}$H$_{12}$,
$\gamma_{\pi^{\ast}\leftarrow\pi}\approx0.55\mum^{-1}$
for dibenzo[cd,\,lm]perylene C$_{26}$H$_{14}$,
and $\gamma_{\pi^{\ast}\leftarrow\pi}\approx0.63\mum^{-1}$
for benzo[g,\,h,\,i]perylene C$_{22}$H$_{12}$).
However,  benzo[a]pyrene
(C$_{20}$H$_{12}$; $\#4$ in Figure~\ref{fig:target}),
a molecule also with a less compact structure,
shows the broadest bump
($\gamma_{\pi^{\ast}\leftarrow\pi}\approx2.9\mum^{-1}$)
in our sample. Triphenylene (C$_{18}$H$_{12}$;
$\#2$ in Figure~\ref{fig:target}), another molecule
with a less compact structure,
also shows a broad bump
($\gamma_{\pi^{\ast}\leftarrow\pi}\approx2.7\mum^{-1}$).
Figure~\ref{fig:specpah} also shows that
``elongated'' compact PAHs like
circumtetracene (C$_{48}$H$_{12}$,
$\#21$ in Figure~\ref{fig:target};
$\gamma_{\pi^{\ast}\leftarrow\pi}\approx2.8\mum^{-1}$),
diphenalenoovalene (C$_{48}$H$_{18}$,
$\#22$ in Figure~\ref{fig:target};
$\gamma_{\pi^{\ast}\leftarrow\pi}\approx2.1\mum^{-1}$),
and circumcircumpyrene (C$_{96}$H$_{26}$,
$\#30$ in Figure~\ref{fig:target};
$\gamma_{\pi^{\ast}\leftarrow\pi}\approx2.1\mum^{-1}$)
exhibit broad bumps.
Therefore, it is clear that a catacondensed
or pericondensed structure {\it alone}
does not determine the bump width.

Apparently, if we carefully select those molecules
characterized by a bump width of
$0.5\simlt\gamma_{\pi^{\ast}\leftarrow\pi}\simlt1.5\mum^{-1}$
(see Figure~\ref{fig:specpah}), a close fit
to the {\it mean} observed bump width of
$\gamma_{\pi^{\ast}\leftarrow\pi}\approx1.0\mum^{-1}$
would be achievable. 
However, the rationale for such a selection
needs to be justified. As mentioned earlier,
the molecules considered here are idealized.
In the ISM, PAHs may include ring defects
(e.g., see Yu \& Nyman 2012),
aliphatic components (e.g., see Yang \& Li 2023a),
oxygen or nitrogen substituents
(Bauschlicher 1998, Hudgins et al.\ 2005,
Mattioda et al.\ 2008),
partial dehydrogenation (Malloci et al.\ 2018)
and sometimes superhydrogenation
(see Bernstein et al.\ 1996, Yang et al.\ 2020)
or deuteration (see Hudgins et al.\ 2004,
Yang \& Li 2023b).
To pin down which exact sets of PAH molecules
are responsible for the observed interstellar
extinction bump, we need to enlarge the sample
by including these non-idealized PAH species.
Nevertheless, it is promising that,
as illustrated in Figure~\ref{fig:specpah},
a mixture of PAH species of certain sizes and
structures apparently are capable of accounting
for the observed bump width, if we assign
appropriate weights to these molecules.
In a subsequent paper, we will model
the 2175$\Angstrom$ extinction bump
observed for a number of representative
interstellar sight lines in terms of PAH mixtures
by, for each sight line, finding the appropriate
weights for these molecules.

Finally, we note that, in addition to graphite and PAHs,
carbon buckyonions composed of spherical concentric
fullerene shells (Chhowalla et al.\ 2003,
Iglesias-Groth et al.\ 2003, Ruiz et al.\ 2005,
Li et al.\ 2008), and T-carbon (Ma et al.\ 2020),
a carbon allotrope formed by substituting
each atom in diamond with a carbon tetrahedron
(Sheng et al.\ 2011), have also been suggested
as a carrier of the interstellar extinction bump.
It would be interesting to see if these candidate
materials are capable of accommodating
the observed characteristics of a stable peak
wavelength at $\simali$2175$\Angstrom$
and a variable width around
$\gamma\approx0.92\mum^{-1}$
in the range of
$0.63\simlt\gamma\simlt1.47\mum^{-1}$
(Valencic et al.\ 2004).



\section{Summary}\label{sec:summary}
We have examined PAHs as a possible carrier of
the mysterious 2175$\Angstrom$ extinction bump
by performing quantum chemical computations
of the absorption spectra of a mixture of
30 compact PAH cations, anions and neutrals
due to electronic transitions.
Our principal results are as follows:
\begin{enumerate}
\item While the absorption spectra of single PAH
          molecules exhibit sharp, individual features,
          for a mixture of many PAH species,
          the individual features of single species
          coalesce to form two broad, prominent
          absorption bands at $\simali$4.4 and
          12$\mum^{-1}$, with the former arising from
          $\pi^{\ast}$\,$\leftarrow$\,$\pi$ electronic transitions
          and the latter arising from transitions
          involving $\sigma$ electrons.
\item The absorption bump resulting from
          $\pi^{\ast}$\,$\leftarrow$\,$\pi$ transitions
          peaks around $\simali$4.4$\mum$, irrespective
          of PAH charge states. The bump position
          is at a somewhat longer wavelength
          than the 2175$\Angstrom$ (4.6$\mum^{-1}$)
          interstellar extinction bump.
          This is probably because
          TD-DFT computations tend to redshift
          the absorption bump compared to
          experimental, gas-phase spectra.
          On the other hand,
          the absorption bump tends to shift to
          longer wavelengths as the PAH size increases.
\item The intensities of the $\pi^{\ast}$\,$\leftarrow$\,$\pi$
          absorption bump do not vary much among
          PAH cations, anions and neutrals.
          This differs from earlier experimental findings
          that, upon ionization, the absorption bump
          substantially weakens. Averaging over all three
          charge states, the intensity (or band strength) is
          $\int_{\pi^{\ast}\,\leftarrow\,\pi}
          \Delta \avgcabsNC\,d\lambda^{-1}
           \approx 2.05\times10^{-13}\cm/{\rm C}$.
           To account for the observed 2175$\Angstrom$
           extinction bump in the diffuse ISM,
           one requires an abundance of
           $\CTOHPAH\approx40\ppm$ to be in PAHs.
\item The widths of the $\pi^{\ast}$\,$\leftarrow$\,$\pi$
          absorption bump of PAH mixtures are considerably
          broader than the interstellar extinction bump
          for otherwise randomly-selected
          and uniformly-weighted PAH mixtures,
          irrespective of the number, sizes, charge states of
          individual molecules and the width assigned to
          each individual electronic transition.
          To account for the observed bump width,
          it appears that a mixture of PAH species of
          specifically-selected sizes and structures
          (i.e., different weights for different molecules)
          are required.
\end{enumerate}

\section*{Acknowledgements}
We thank B.T.~Draine and the anonymous referee
for valuable suggestions.
QL and XJY are supported in part by
NSFC 12122302 and 11873041.
AL is supported in part by NASA grants
80NSSC19K0572 and 80NSSC19K0701.

\section*{Data Availability}
The data underlying this article will be shared
on reasonable request to the corresponding authors.


%

%

\begin{table*}
\caption[]{\footnotesize
  Parameters for fitting the mean absorption spectra
  of PAH cations (PAH$^{+}$), anions (PAH$^{-}$),
  and neutrals with a Drude profile and a Fano profile.}
\label{tab:para}
\centering
{\scriptsize
\begin{tabular}{lcccccccc}
\noalign{\smallskip} \hline \hline \noalign{\smallskip}
Species &
$a_{1} $ &
$x_{0,1}$&
$\gamma _{1}$ &
$a_ {2} $ &
$x_{0,2} $ &
$\gamma _{2}$ &
$\mathit{q}$ & \\
&
($\times10^{-10}$) &
($\mum^{-1}$) &
($\mum^{-1}$) &
($\times10^{-18}\cm^2/{\rm C}$) &
($\mum^{-1}$) &
($\mum^{-1}$) &
 & \\
\noalign{\smallskip} \hline \noalign{\smallskip}
$\mathrm{PAH}^{+}$  &
20.15\,$\pm$\,2.12   &
4.45\,$\pm$\,0.03  &
1.52\,$\pm$\,0.09  &
2.16\,$\pm$\,0.11  &
12.13\,$\pm$\,0.05  &
7.00\,$\pm$\,0.10  &
3.13\,$\pm$\,0.09  & \\
$\mathrm{PAH}^{-}$  &
21.25\,$\pm$\,1.93   &
4.35\,$\pm$\,0.03  &
1.56\,$\pm$\,0.08  &
1.67\,$\pm$\,0.08  &
12.27\,$\pm$\,0.05  &
7.65\,$\pm$\,0.09  &
3.57\,$\pm$\,0.10  & \\
$\mathrm{PAH}^{0}$  &
21.11\,$\pm$\,1.59   &
4.41\,$\pm$\,0.02  &
1.56\,$\pm$\,0.07  &
1.88\,$\pm$\,0.07  &
12.26\,$\pm$\,0.04  &
7.39\,$\pm$\,0.07  &
3.35\,$\pm$\,0.07  & \\
Average  &
21.20\,$\pm$\,1.72   &
4.40\,$\pm$\,0.02  &
1.56\,$\pm$\,0.07  &
1.90\,$\pm$\,0.08  &
12.22\,$\pm$\,0.05  &
7.34\,$\pm$\,0.08  &
3.34\,$\pm$\,0.08  & \\
\hline
\noalign{\smallskip}
\noalign{\smallskip} \noalign{\smallskip}
\end{tabular}
}
\end{table*}

\clearpage

\clearpage

\clearpage

\clearpage

\clearpage

%
\end{document}